\documentstyle[epsfig]{aipproc}

\begin{document}

\title{NMR studies of the original magnetic properties of the cuprates:
effect of
impurities and defects}
\author{H.\ Alloul, J.\ Bobroff, A.\ Mahajan, P.\ Mendels and Y.\ Yoshinari.}

\address{Laboratoire de Physique des Solides, UMR 8502 CNRS,\\
Universit\'{e}
Paris-Sud
91405, Orsay, France.}

\maketitle

\begin{abstract}
Substitutional impurities in the CuO$_{2}$ planes of the cuprates allow us
to probe the electronic properties of the host material.\ The pseudo-gap in
the underdoped regime is unmodified far from the impurities even though $%
T_{c}$ is greatly reduced. The spin polarisation induced by magnetic
impurities has an oscillatory behaviour reflecting the existing AF\
correlations between the Cu spins.\ Its influence on the NMR spectra opens a
way to determine the ${\bf q}$ dependence of the static spin susceptibility
and the $T$ dependence of the AF correlation length. NMR\ measurements
demonstrate that non-magnetic impurities such as Zn induce a local moment
behaviour on the neighbouring Cu sites.\ This magnetism revealed by
spin-less sites can be understood on theoretical grounds in the case of
undoped quantum spin systems, while here the carriers greatly complicate the
situation.\ Susceptibility data show that the magnitude of the local moment
decreases with increasing hole doping.\ This experimental evidence
directly reflects the influence of AF correlations and the interference
between the carriers and the Cu hole spins in the cuprates. The anomalously
large scattering of the carriers on spinless defects is another indication
of the originality of the electronic properties of the cuprates, which
apparently extends even to the overdoped regime.
\end{abstract}

\section{Introduction: magnetic properties of pure materials}

It is now experimentally well established that the CuO$_{2}$ planes display
anomalous magnetic properties in the metallic normal state of the cuprates,
at least in the underdoped and optimally doped states. The occurrence of
magnetic correlations was first shown by the existence of an enhanced
non-Korringa nuclear spin relaxation rate $1/T_{1}$ on $^{63}$Cu and not on $%
^{17}$O\ and $^{89}$Y \cite{takigawa,alloulohno}. \ In the recent past,
considerable interest has been focused on the pseudo-gap in the excitation
spectrum of the cuprates.\ It was detected first in microscopic NMR
measurements of the susceptibility $\chi _{p}$ of the CuO$_{2}$ planes \cite
{alloulohno}, which exhibit a large reduction in the homogeneous ${\bf q}=0$
excitations at low $T$ in underdoped materials, as shown in Fig.1. Similar
low $T$ reductions of the imaginary part of the susceptibility at the AF
wave vector ${\bf q}=(\pi ,\pi )$ were observed in $^{63}$Cu $1/T_{1}T$ data
and inelastic neutron scattering experiments \cite{berthier}. Presently, the
pseudo-gap of the underdoped high-$T_{c}$ superconductors is also studied by
many techniques such as transport and angle resolved photoemission, which
yields its ${\bf k}$ dependence. Various explanations are proposed for the
pseudo-gaps which are believed to be essential features of the physics of
the normal state (and perhaps the superconducting state) of the cuprates.

\smallskip 
\begin{figure}[tbp]
\begin{center}
\epsfig{file=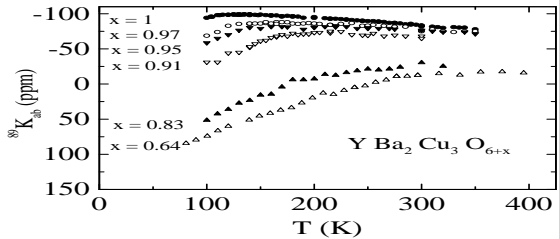,figure=fig1.eps,height=7cm,width=8.cm,angle=0,}
\end{center}
\caption{The $^{89}$Y\ NMR\ shift data evidence the occurrence of pseudo-gaps
from a large decrease of the susceptibility $\chi _{p}$ of the CuO$_{2}$\
planes, in YBCO$_{6+x}$.\ The temperature at which the pseudo-gap begins to
open, which corresponds to the $\chi _{p}$ maximum, increases markedly with
decreasing $x,$ that is decreasing hole doping (theoriginal data
\protect\cite{alloulohno} are less accurate that those displayed here).}
\end{figure}

Atomic substitutions in the planar Cu site have naturally been found the
most detrimental to superconductivity. This has in parallel triggered a
large effort, particularly in our research group, to use such impurities to
reveal the original normal-state magnetic properties of the cuprates.\ We
shall see that NMR,\ as a local magnetic probe, is an essential tool which
lends weight to this approach.\ In section II we present the effect of
impurities on the phase diagram and the pseudo-gaps.\ We will distinguish
the average effect of impurities on the physical properties far from the
impurity site from the local magnetic perturbations.\ The study of the
distribution of the spin polarisation induced by magnetic impurities is
shown in sec.\ III to be a direct probe of the non-local magnetic response $%
\chi ^{\prime }({\bf r})$ of the pure system, a quantity which is hard to
access by other experimental approaches.\ In section IV\ we consider the
case of non-magnetic impurities like Zn ($3d^{10}$) which, upon substitution
on the Cu site of the CuO$_{2}$ plane, strongly decreases the
superconducting transition temperature $T_{c}$. It has been anticipated \cite
{fink}, and subsequently shown experimentally\cite{alloul2}, that, although
Zn itself is non-magnetic, it induces a modification of the magnetic
properties of the correlated spin system of the CuO$_{2}$ planes. We shall
recall how $^{89}$Y NMR demonstrated \cite{mahajan} that local magnetic
moments are induced on the Cu near neighbour ($n.n.$) of the Zn substituent
in the CuO$_{2}$ plane. Experiments on La$_{2-x}$Sr$_{x}$CuO$_{4}$ have
confirmed \cite{ishida2} that the occurrence of local moments induced by
non-magnetic impurities on the Cu sites is a general property of cuprates.
Recent measurements of the variation with hole doping of the effective
magnetic moment associated with non magnetic impurities will be reviewed.
Finally, we shall discuss briefly in sec.V the influence of impurities on
transport properties.

\section{Impurities and phase diagrams}

\smallskip The main effect of impurities is to depress
superconductivity, usually much faster in the underdoped regime than
in the optimally doped case.\ Similarly the increase of resistivity is
larger for underdoped materials.\ This results in a shift of the
Insulator-Metal transition towards higher hole concentration in
presence of impurities.\ So what happens then to the crossover
pseudo-gap lines?\ There is a controversy which has arisen because of
the differences between macroscopic and microscopic measurements of
the pseudo-gap.\ NMR\ has the advantage that the sites in the vicinity
of the impurity usually display well shifted resonance lines.\
Therefore, the main NMR line corresponds to sites far from the
impurity.  Its broadening, to be studied in section III, is associated
with the oscillatory induced polarisation of the host.  Its position
measures then the average $\chi _{p}$ far from the impurities which
reflects their influence on the homogeneous
magnetic properties. The $T$ dependence of the shifts $\Delta K(T)$ of the $%
^{89}$Y or $^{17}$O NMR\ mainlines \cite{alloul2,mahajan,bobroff} are found
to be unmodified by impurity substitutions, as can be seen in Fig.\ 2.

This demonstrates that, contrary to the Metal-Insulator transition, the {\it %
pseudo-gap is unaffected} by impurity substitutions at large distance from
the impurities. Incidentally these data also ensure that the hole doping is
not significantly modified.\ Conversely if hole doping is changed, i.e. by
Pr substitution on the Y site \cite{macfarlane}, NMR shift data can be used
to estimate its variation by comparison with calibrated curves for $\chi
_{p} $ in the pure material (Fig\ 1).\

\smallskip 
\begin{figure}[t]
\begin{center}
\epsfig{file=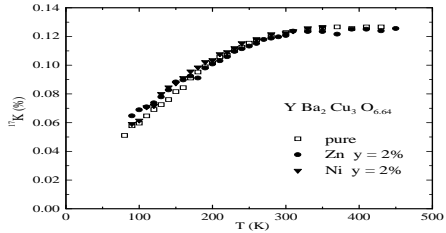,figure=fig2.eps,height=7cm,width=7.5cm,angle=0,}
\end{center}
\par
\caption{ The spin susceptibility $\chi _{p}$ and therefore the temperature $%
T^{*}$ of the $^{17}$O NMR shift maximum, characteristic of the opening of
the pseudo-gap, are not modified upon Zn or Ni substitution, while $T_{c}$
is significantly reduced (from 60K to $\approx $20K for 2\% Zn) (similar
data were taken first \protect\cite{alloul2} on $^{89}$Y).}
\end{figure}
\

Other experiments, which very often probe the macroscopic behaviour of the
sample, have sometimes been interpreted differently, as they do not directly
distinguish local from large distance properties. For instance, it was
initially suggested \cite{kakurai} on the basis of neutron scattering
experiments, that the pseudo-gap vanishes at ${\bf q}=(\pi ,\pi )$ upon Zn
substitution. However, the careful neutron data of Sidis {\it et al.} \cite
{sidis} indicate that the opening of the pseudo-gap still occurs at the same
$T^{*}$ although new states appear in the pseudo-gap. Such states may be
associated with local magnetic modifications induced around the Zn, which
will be described in section IV.\ In any case, far from the impurities the
pseudo-gap is quite robust upon impurity substitution. These results are
quite natural if the pseudo-gaps are only associated with the occurrence of
AF correlation effects.\ However, in the scenario in which the pseudo-gaps
are associated with the formation of local pairs below $T^{*}$, they
indicate that impurities do not prevent the formation of local pairs except
possibly in their vicinity.

\section{\protect\smallskip Extended response to a local magnetic excitation}

\smallskip 
\begin{figure}[tbp]
\begin{center}
\epsfig{file=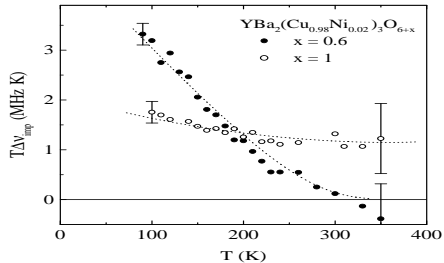,figure=fig3.eps,height=7cm,width=7.5cm,angle=0,}
\end{center}
\par
\caption{The $^{17}$O NMR\ broadening $\Delta \nu _{imp\text{ }}$ multiplied
by $T$ is plotted versus $T$ for Ni substituted YBCO\ samples \protect\cite
{bobroff}.\ The variation of this quantity is related to that of the
susceptibility $\chi ^{\prime}$(${\bf q}$), which is peaked near the AF wave
vector.\ The dotted lines are guides to the eye.}
\end{figure}

\smallskip In noble metals hosts, any local charge perturbation is known to
induce long distance charge density oscillations (also called Friedel
oscillations). Similarly a local magnetic moment induces a long distance
oscillatory spin polarisation (RKKY) which has an amplitude which scales
with the magnetization of the local moment and with its coupling $J_{ex}$
with the conduction electrons. This oscillatory spin polarisation gives a
contribution to the NMR shift of the nuclei which decreases with increasing
distance from the impurity. In very dilute samples, if the experimental
sensitivity is sufficient, the resonances of the different shells of
neighbours of the impurity can be resolved \cite{alloul1}. These resonances
merge together if the impurity concentration is too large, resulting in a
net broadening $\Delta \nu _{imp\text{ }}$of the host nuclear resonance. For
some impurities, like rare earths substituted on the Y site, the
hybridisation and therefore the exchange coupling $J_{ex}$ are extremely
weak.\ The plane nuclear spins only sense the moment through its dipolar
field. But for moments located in the planes such as Ni, the induced spin
polarisation dominates, especially for the $^{17}$O and $^{63}$Cu nuclei.
The large $^{17}$O NMR\ broadening induced by Ni has been therefore studied
in great detail by Bobroff {\it et al }\cite{bobroff}. It has been found
that in underdoped YBCO, the linewidth increases much faster than $1/T$ at
low temperature, contrary to what one might expect in a non-correlated
metallic host (Fig.3). This fast increase is a signature of the anomalous
magnetic response of the host which displays a peak in $\chi ^{\prime }({\bf %
q})$ near the AF wavevector ($\pi $, $\pi $), which can be characterized by
a correlation length $\xi $. Therefore the $T$ variation of the NMR spectra
yields a method to study the $T$ dependence of $\chi ^{\prime }({\bf q})$
and $\xi $. The analysis of such data depends on the phenomenological shape
given to the ${\bf q}$ dependence of $\chi ^{\prime }({\bf q}),\;$especially
as the O\ site probes $\chi ^{\prime }({\bf r})$\ on its two neighbours, and
therefore is governed somewhat by the gradient of $\left| \chi ^{\prime }(%
{\bf r})\right| $. Assuming a Gaussian shape for $\chi ^{\prime }({\bf q})$,
it is found that the linewidth is nearly insensitive to the $T$ dependence
of $\xi $, in contrast to the case of a Lorentzian shape \cite{slichter}.\
The experiment on a single nuclear site does not by itself allow deduction
of $\xi (T)$.\ However comparison with spin-spin relaxation data, which also
measures an integral quantity involving $\chi ^{\prime }({\bf q})$ yields
some complementary information. With a Gaussian shape we find that $\xi $
increases with $T$ while for a Lorentzian, it would decrease with increasing
$T$.\ Similarly a comparison of the respective broadenings of the $^{17}$O, $%
^{89}$Y and $^{63}$Cu spectra, which probe differently $\chi ^{\prime }({\bf %
r})$, lead us to conclude that the Lorentzian model is somewhat better,
implying that $\xi $ increases at low $T$ as one might expect \cite{bobroff2}%
.\ More accurate studies of the shape of the spectra as well as their
concentration dependence are required to arrive at quantitative conclusions
on the variation of $\xi (T)$. We should point out here that, for YBCO$_{7}$%
, the $^{17}$O width varies roughly like $1/T$, while the Ni moment still
displays a Curie law.\ This indicates that the $T$ dependence of $\chi
^{\prime }({\bf q})$ is not as large in optimally doped systems.\ In such
cases the detailed shape of $\chi ^{\prime }({\bf q})$ could not be analysed
up to now.

\section{\protect\smallskip Local magnetism induced by non-magnetic Zn}

Surprisingly it has been found that, as for Ni substitution, a $1/T$\
broadening of the $^{89}$Y\ line occurs \cite{alloul2} in YBCO$_{7}$:Zn,
even though Zn is expected to be in a non-magnetic 3d$^{10}$ state. This was
the first experimental evidence{\it \ for the occurrence of ``local moment
like'' behaviour induced by Zn}. A more refined picture of the response of
the host to a non-magnetic substituent has been obtained in the case of
underdoped YBCO$_{6.64}$, as distinct resonances of $^{89}$Y were observed
and could be attributed to Y $n.n.$ sites of the substituted Zn \cite
{mahajan}. The measured Curie-like contribution to the NMR shift of the
first $n.n.$ line (Fig.4), and the shortening of its T$_{1}$ at low-$T$ are
striking evidence which justify the denomination ``local moment'', that we
have been using throughout \footnote{%
The validity of these observations has been periodically put into question,
for instance as similar $n.n.$ resonances were not detected\cite{janossy} in
ESR experiments on Gd (substituted on Y). From the T$_{1}$ data for $^{89}$
Y\ NMR, we have shown that the large expected relaxation rate for Gd
corresponds to a significant line broadening of the Gd ESR $n.n.$ lines
which prohibits their detection \cite{nmahajan}. It has also been
conjectured that substitution of Zn on Cu and Ca on Y yield similar disorder
effects on the NMR \cite{tallon}.\ This is not true as the line broadening
does not exhibit a Curie like T variation for Ca substituted samples.}.

\smallskip 
\begin{figure}[tbp]
\begin{center}
\epsfig{file=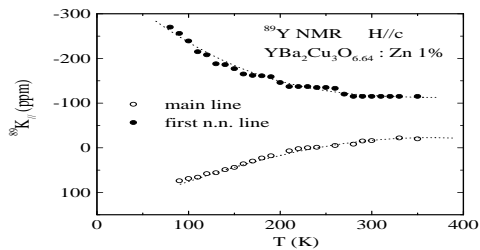,figure=fig4.eps,height=7cm,width=7.5cm,angle=0,}
\end{center}
\caption{ The $n.n.$ $^{89}$Y\ NMR
shift obtained in YBCO$_{6.64}$:Zn displays a paramagnetic Curie like
component in contrast to the decreasing (pseudo-gap) susceptibility of the
host material \protect\cite{mahajan}.}
\end{figure}

The Zn induced local moments are quite clearly located in the vicinity of
the Zn. As is shown by analysis of the $n.n.$ $^{89}$Y NMR intensity data,
the Zn substitutes essentially on the plane copper site.\ Therefore, the Zn
contribution $\chi _{c}$ to the macroscopic susceptibility could be inferred
from SQUID data taken on samples free of parasitic impurity phases \cite
{mendels,mendels1}. The hyperfine couplings deduced from the comparison of
the $^{89}$Y\ NMR\ shift data to $\chi _{c}$ have the correct order of
magnitude to demonstrate that the local moment resides mainly on the Cu $%
n.n. $ to the Zn. Assuming that they are not modified with respect to pure
YBCO, the data can be analysed consistently with {\it a locally AF state
extending over a few lattice sites}. This might also explain the existence
of a line corresponding to Y\ second $n.n.$ to the Zn \cite{nmahajan}.

In the superconducting state, $^{63}$Cu NQR relaxation \cite{ishida1} and
M\"{o}ssbauer experiments \cite{hodges} indicate the existence of states in
the gap. In neutron scattering experiments \cite{sidis}, the local states
induced by the Zn, both in the pseudo-gap and in the spin-gap detected below
$T_{c}$, are found at the ($\pi $, $\pi $) scattering vector, and correspond
to a real state extension of about 7\AA .{\it \ } These thus constitute
direct evidence for the persistence of AF correlations in the vicinity of
the impurities \cite{comment}.

It is clear that the observed local moment behaviour is original inasmuch as
it is the {\it magnetic response of the correlated electron system to the
presence of a spinless site,} as has been proposed from various theoretical
arguments \cite{fink,nagaosa,poilblanc,khaliullin,nagaosalee}. As complete
understanding of the magnetic properties of pure cuprates is far from being
achieved, it is no surprise that present theoretical descriptions of
impurity induced magnetism are rather crude, and for example, do not address
its microscopic extent. Our results also fit well in the context of recent
theoretical work on undoped quantum spin systems. For instance Martins \cite
{dagotto} predicts static local moments induced by doping $S=1/2$ Heisenberg
AF chains or ladders with non-magnetic impurities. NMR experiments on the $%
S=1/2$ Heisenberg chain system Sr$_{2}$CuO$_{3}$ are consistent with the
prediction of an induced local moment with a large spatial extent along the
chain \cite{takigawa2}. In this undoped insulating quantum liquid, the
response is purely magnetic. Since AF correlations persist in the metallic
cuprates, the appearance of a local moment near the Zn might be anticipated,
but its properties could depend strongly on the density of charge carriers.\
Magnetisation data by Mendels {\it et al.} \cite{mendels,mendels1}, shown in
Fig. 5, demonstrate that the Curie constant decreases steadily from YBCO$%
_{6.64}$:Zn to YBCO$_{7}$:Zn. However existing experiments do not, at
present, distinguish the respective roles of the AF correlation length and
screening by the conduction holes in defining the local moment magnitude and
spatial extent.

\smallskip
\begin{figure}
\begin{center}
\epsfig{
figure=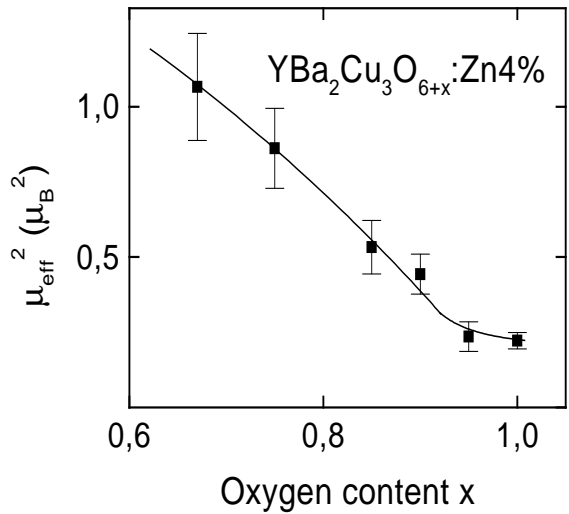,%
height=7cm,%
width=7.5cm,%
angle=0,%
}
\end{center}
\caption{The Curie constant for the local moment induced by Zn in
YBCO$_{6+x}$, as measured by macroscopic magnetic susceptibility
measurements is found to decrease markedly with hole doping
\protect\cite{mendels1}}
\end{figure}

\smallskip In the slightly overdoped YBCO$_{7}$, the occurrence of a local
moment was confirmed from $^{17}$O NMR linewidth data \cite{YY}.\ The fact
that we could not resolve the $^{89}$Y$\;n.n.$ signal in YBCO$_{7}$ is
consistent with the weak magnitude found for the Curie-like contribution to
the local susceptibility. Furthermore, Ishida {\it et al \thinspace }\cite
{ishida2}{\it \ }showed that our observation extends to another cuprate
family, as non-magnetic Al exhibits a local moment behaviour in optimally
doped La$_{2-x}$Sr$_{x}$CuO$_{4}$. A local signature of this fact was found
in the shift of the $^{27}$Al NMR which exhibits a Curie-Weiss T dependence
of $\chi _{c}$, with a sizable Weiss temperature ($\theta $ $\approx $ $50K$%
). It is not very clear from these data whether such a high value of $\theta
$ corresponds to a genuine single impurity effect or if itvaries with Al
content, thereby revealing a strong coupling between the local moments. By
analogy with our results this observation is supposed to result from a local
moment residing on the $n.n.$ copper orbitals, which are coupled to the $%
^{27}$Al nuclear spin via transferred hyperfine couplings.

\section{Magnetism and transport properties}

Let us now consider briefly the influence of impurities on{\ {\it transport
properties}. } Most analyses of the resistivity data \cite{ong,mizuhashi}
suggest a large magnitude for the Zn scattering, not far from the unitary
limit. Such results are generally observed for most local defects induced in
the CuO$_{2}$ planes, such as irradiation defects \cite{legris}. It has
occasionally been assumed \cite{tallon2} that strong scattering is due to
potential scattering. In the present case, for which no charge difference
occurs between Zn$^{2+}$ and Cu$^{2+}$, the scattering cannot be associated
with charge difference, but rather is due indirectly to the fact that Zn is
a spin-less defect. So, as for Kondo impurities in normal metal hosts,
unitary scattering is associated with a magnetic effect. These remarks are
of course included in most analyses of impurity scattering done in the
strong correlation approaches \cite
{fink,nagaosa,poilblanc,khaliullin,nagaosalee}. Different behaviour of
spinons and holons is at the root of the anomalous impurity scattering in
such theories. However none is at present sufficiently advanced to provide
quantitative results which could be compared to experimental data.

As for the reduction of T$_{c}$ induced by ``non-magnetic'' impurities, the
situation has evolved since our first report \cite{mahajan}, since the d
wave symmetry of the order parameter is now well established in most
cuprates. In this case, any type of impurity scattering depresses $T_{c}$.
This is exemplified by recent resistivity measurements performed on electron
irradiated samples \cite{albenque}.\ It it is found that a universal law
applies between the respective variations $\Delta T_{c}$ and $\Delta \rho $
of the superconducting temperature and of the resistivity induced by
defects.\ The most remarkable feature detected in these experiments is that
the universal relation extends to the overdoped regime, suggesting that the
d wave symmetry of the order parameter is valid for the entire phase
diagram, even in the doping range for which Fermi liquid behaviour seems to
apply.\

\section{Conclusions}

\smallskip The studies presented above have allowed us to demonstrate that
valuable information on the magnetic properties of the cuprates are obtained
from the local study of their response to substitutional impurities in the
CuO$_{2}$ planes. It could be anticipated, by analogy with RKKY effects in
simple metals, that the large distance modifications of the host properties
would directly reflect the non-local magnetic response of the host. This is
indeed apparent in the NMR studies of the modifications of the central $%
^{89} $Y\ or $^{17}$O NMR lines, which give us some insight on the ${\bf q}$
and $T $\ dependence of the static susceptibility, i.e.\ of the AF\
correlation length.\ These studies also show that the pseudo-gap is
unaffected by impurities and is therefore either a purely magnetic effect
associated with AF fluctuations, or due to a local pairing which is not
disrupted by impurities in contrast with the macroscopic superconducting
condensate.

More surprising are the actual properties associated with spin-less
impurities.\ The existence of magnetic correlations are responsible for the
occurrence of local moment behaviour induced by non-magnetic impurities on
the neighbouring Cu sites.\ While one might expect that the spin 1/2 moment
which is released by such a defect in a magnetic quantum spin system should
extend on a distance comparable to the AF correlation length, the measured
decrease of the effective moment with increasing hole content is more likely
to be associated with an interaction with charge carriers.\ As transport
properties indicate that isovalent impurities produce a large scattering of
the carriers, one can anticipate that this is due to the occurrence of a
resonant state.\ But the actual Curie dependence of the susceptibility would
not be expected then to extend to low T. However, the occurrence of
superconductivity limits somewhat our experimental capability to investigate
the magnetic properties of the metallic ground state.\ From SQUID data \cite
{mendels1} it was concluded that in underdoped YBCO$_{6.6}$:Zn4\%, for which
$T_{c}$ is reduced to zero, the susceptibility follows a Curie-Weiss law
down to 10K, with $\theta $ $\approx 4K.$ Does this correspond to a
Kondo like energy scale characteristic of the width of the impurity resonant
state? Such a Kondo-like effect is a candidate mechanism \cite{nagaosalee}
for the reduction of the magnitude of the local moment in YBCO$_{7}$:Zn.
However in these systems, one does not a priori expect the physical
properties to mimic those obtained for Kondo effect in noble metals in a
classical sd exchange model.\ Such difficulties have already been pointed
out by Hirschfeld \cite{hirschfeld} in view of our preliminary data \cite
{mahajan}. Obviously further efforts are required to complete this approach
both experimentally and theoretically in the context of a correlated
electron system.

We should like to acknowledge A.\ MacFarlane for helpful discussions
and careful reading of the manuscript.\


\smallskip

\begin{thebibliography}{99}
\bibitem{takigawa}  M.~Takigawa {\it et al}, Phys. Rev. Lett. {\bf 63}, 1865
(1989); Phys.~Rev.~{\bf B} {\bf 43}, 247 (1991).

\bibitem{alloulohno}  H.~Alloul, T.~Ohno, and P.~Mendels, Phys.~Rev.~Lett.~%
{\bf 63}, 1700 (1989).

\bibitem{berthier}  C.Berthier, M.\ H.\ Julien, M.\ Horvatic and Y.\
Berthier, Appl. Magn. Reson. {\bf 3}, 449 (1992).

\bibitem{fink}  A.~M.~Finkelstein, V.~E. Kataev, E.~F.~ Kukovitskii, and
G.~B.~Teitelbaum, Physica C {\bf 168}, 370 (1990).

\bibitem{alloul2}  H.~Alloul {\it et al,} Phys.~Rev.~Lett. {\bf 67}, 3140
(1991).

\bibitem{mahajan}  A.~V.~Mahajan, H.~Alloul, G.~Collin, J.~F.~Marucco,
Phys.~Rev.~Lett.~{\bf \ 72}, 3100 (1994).

\bibitem{ishida2}  K.~Ishida {\it et al,} Phys.~Rev.~Lett.~{\bf 76}, 531
(1996).

\bibitem{bobroff}  J.~Bobroff {\it et al,} Phys. Rev. Lett. {\bf 79}, 2117
(1997).

\bibitem{macfarlane}  A.\ MacFarlane {\it et al}, to be published.

\bibitem{kakurai}  K.~Kakurai {\it et al,} Phys.~Rev.~{\bf B} {\bf 48}, 3485
(1993).

\bibitem{sidis}  P.~Bourges {\it et al,} Journal of Physics {\bf 46}, 1155
(1996); P.~ Bourges, {\it et al,} J. Phys. Chem. Solids {\bf 56}, 1937
(1995); Y.~ Sidis {\it et al.}, Int. J. of Modern Phys. B, to be published.

\bibitem{alloul1}  J.~B.~Boyce and C.~P.~Slichter, Phys.~Rev.~Lett.~ {\bf 32}%
, 61 (1974); H.~Alloul, F.~Hippert, and H.~Ishii, J.~Phys.~F {\bf 4}, 725
(1979) and references therein.

\bibitem{slichter}  D.\ K.\ Morr, J.\ Schmalian, R.\ Stern and C.\ P.\
Slichter, Phys. Rev.\ Lett. {\bf 80}, 3662 (1998); J.~Bobroff {\it et al,}
Phys. Rev. Lett. {\bf 80}, 3663 (1998).

\bibitem{bobroff2}  J.\ Bobroff, thesis (unpublished).

\bibitem{janossy}  A.~Janossy, J.~R.~Cooper, L.~C.~Brunel, and
A.~Carrington, Phys.~Rev.~{\bf B} {\bf 50}, 3442 (1994).

\bibitem{nmahajan}  A.~V.~Mahajan, H.~Alloul, G.~Collin, J.~F.~Marucco, to
be published.

\bibitem{tallon}  G.\ V.\ M.\ Williams and J.\ L.\ Tallon, Phys.\ Rev.\ {\bf %
B 57}, 10984 (1998).

\bibitem{mendels}  P.~Mendels {\it et al,} Phys.~Rev.~{\bf B 49}, 10035
(1994).

\bibitem{mendels1}  P.~Mendels {\it et al}, to be published in Europhysics
Letters.

\bibitem{ishida1}  K.~Ishida {\it et al}, J.~Phys.~Soc.~Japan {\bf 62}, 2803
(1993).

\bibitem{hodges}  J.~A.~Hodges, P.~Bonville, P.~Imbert, and
A.~Pinatel-Philippot, Physica C {\bf 323}, 246 (1995).

\bibitem{comment}  H.~Alloul, J.~Bobroff, and P.~Mendels, Phys. Rev. Lett.
{\bf 78}, 2494 (1997).

\bibitem{nagaosa}  N.~Nagaosa and T.~K.~Ng, Phys.~Rev.~{\bf B} {\bf 51},
15588 (1995).

\bibitem{poilblanc}  D.~Poilblanc, D.~J.~Scalapino, and Hanke, Phys. Rev.
Lett. {\bf 72}, 884 (1994); see also Phys. Rev. {\bf B} {\bf 50}, 13020
(1994).

\bibitem{khaliullin}  G.~Khaliullin, R.~Killian, S.~Krivenko, and P.~Fulde,
Physica C {\bf 282-287}, 1749 (1997).

\bibitem{nagaosalee}  A.~Nagaosa and P.~Lee, Phys. Rev. Lett. {\bf 79}, 3755
(1997).

\bibitem{dagotto}  G.~B.~Martins, M.~Laukamp, J.~ Riera, and E.~Dagotto,
Phys. Rev. Lett. {\bf 78}, 3563 (1997).

\bibitem{takigawa2}  M.~Takigawa, N.~Motoyama, H.~Eisaki, and S.~Uchida,
Phys. Rev. {\bf B} {\bf 55}, 14129 (1997).

\bibitem{YY}  Y. Yoshinari, J. Bobroff {\it et al} (unpublished).

\bibitem{ong}  T.~R.~Chien, Z.~Z.~Wang, and N.~P.~Ong, Phys.~Rev.~Lett.~{\bf %
67}, 2088 (1991).

\bibitem{mizuhashi}  K.~Mizuhashi, K.~Takenaka, Y.~Fukuzumi, and S.~Uchida,
Phys.~Rev.~{\bf B} {\bf 52}, R3884 (1995).

\bibitem{legris}  A.\ Legris, F.Rullier-Albenque, E.\ Radeva and P.\ Lejay,
J. de Physique (France) I {\bf 3}, 1605 (1993).

\bibitem{tallon2}  C.\ Bernhardt {\it et al}, Phys.\ Rev.\ Lett. {\bf 77},
2304 (1996).

\bibitem{albenque}  F. Rullier-Albenque {\it et al }(to be published).

\bibitem{hirschfeld}  L.~S.~Borkowski and P.~J.~Hirschfeld, Phys. Rev. {\bf B%
} {\bf 49}, 15404 (1994).
\end{thebibliography}
\end{document}